# The precise time-dependent solution of the Fokker-Planck equation with anomalous diffusion

Guo Ran,  Du Jiulin

*Department of Physics, School of Science, Tianjin University, Tianjin 300072, China*

**Abstract** We study the time behavior of the Fokker-Planck equation in Zwanzig's rule (the backward-Ito's rule) based on the Langevin equation of Brownian motion with an anomalous diffusion in a complex medium. The diffusion coefficient is a function in momentum space and follows a generalized fluctuation-dissipation relation. We obtain the precise time-dependent analytical solution of the Fokker-Planck equation and at long time the solution approaches to a stationary power-law distribution in nonextensive statistics. As a test, numerically we have demonstrated the accuracy and validity of the time-dependent solution.

## 1. Introduction

The Fokker-Planck (FP) equation was first applied to the Brownian motion problem [1]. With the equation of motion of a Brownian particle, Langevin equation, and the corresponding FP equation, the probability distribution to find the particle in a given region may be determined by solving the equation. The simplest situation of the Brownian motion is a Brownian particle moving in the medium with friction constant $\gamma$ and diffusion constant $D$, and the link between the two constants is $D=\gamma kT$, known as the fluctuation-dissipation relation (FDR) [2]. In such a situation the Langevin equation and the FP equation are both linear and the solutions (stationary and time-dependent) are Gaussian distributions or Maxwell-Boltzmann (MB) distributions. But for a general situation when a Brownian particle is moving in a complex medium in which the friction and diffusion coefficient can depend on the variables, the Langevin equation is nonlinear and then solving the corresponding FP equation becomes very complicated. In fact, not much has been known in general about the long-time steady-state solution of an arbitrary FP equation. Only in some special cases if a FDR can be invoked, a steady-state solution is found.

The Brownian motion characterized as a pure diffusion process has a probability distribution that is Gaussian at all times and obeys the Einstein relation at long time, the mean-square displacement $<(\Delta x)^2>=2Dt$, where $D$ is a constant, which is called normal diffusion. Anomalous diffusion is random motion having $<(\Delta x)^2> \sim t^\nu$ with $\nu \neq 1$ and therefore there is no constant diffusion coefficient ($D$ may be space/velocity dependent [3-9]) and the associated probability distribution is non-Gaussian or non-MB/power-law distributions [10-17]. Many nonlinear FP equations which appear to be some "fractal structure" are frequently constructed to describe the systems which behave anomalous diffusion [18-23]. It is interesting that they found the steady-state solution following a power-law $q$-distribution in nonextensive statistics [24]. However, these fractal FP equations are all "nonstandard" and due to the lack of physically corresponding Langevin equation, the dynamical origins of the power-law distribution and the physical mechanism that leads to such a distribution are unknown.

Non-Gaussian or non-MB/power-law distributions have been noted prevalently in physical, chemical, biological and even social systems. In recent years, theoretical and experimental researches of these distributions have attracted great attention in the various fields of science, such as astronomy and astrophysics [25-29], plasmas and space physics [10,14,30-34], and reaction rate theory in chemistry [35-39] etc. In terms of the above studies, the power-law distributions often link to the complex systems involving long-range interactions, inhomogeneity and non-equilibrium dissipation processes. Information about the dynamical origins of these anomalous distributions is important for the understanding of many different processes in complex systems. This problem may be seeking a solution from the standard FP equation based on the Langevin equation for the dynamics of Brownian motion [16]. We have studied a general position-momentum Brownian motion in an inhomogeneous medium and with a multiplicative noise. The diffusion coefficient and friction coefficient can be position/momentum dependent for a Brownian particle moving in complex medium. By invoking a generalized FDR one could seek the steady-state solutions from the standard FP equations in both Ito's, Stratonovich's and Zwanzig's (or the backward Ito's) rules, where many different forms of power-law distributions were found [16,17]. Besides the steady-state solutions, the time-dependent solutions of the FP equations are also important for us to understand the dynamical evolution of the probability distributions in complex systems. However it is not easy to solve a general multivariable FP equation. In this work, we will try to find the time-dependent solution of the standard FP equation with an anomalous diffusion in momentum space.

The paper is organized as follows. In section 2, we briefly describe the standard FP equation based the Langevin equation for Brownian motion in a complex medium, the generalized FDR and its associated power-law distribution. In section 3, we solve the time-dependent FP equation with an anomalous diffusion in momentum space. The precise time-dependent analytical solution will be given. In section 4, numerical studies are made to examine the accuracy and validity of the analytical solution, including a general test and the application to the Ornstein-Uhlenbeck process. Finally in section 5, we give the conclusion.

## 2. The Fokker-Planck equation and power-law distribution

We consider a Brownian particle, with the mass $m$, moving in a medium with a friction coefficient $\gamma$ as well as a noise $\eta$, and under a potential field $V(x)$. In the simplest case, the friction and diffusion coefficients may be regarded as constant approximately. But this is not always true. For example, the plasma immerged in a superthermal radiation field would lead to a multiplicative stochastic process in the velocity-space diffusion and therefore the friction and diffusion coefficients are both a function of the velocity [10]. For generality, when the Brownian particle moves in an inhomogeneous complex medium, the friction coefficient is a function of the variables $x$ and $p$, i.e. $\gamma=\gamma(x,p)$, and the noise is multiplicative and position/momentum-dependent, i.e. $\eta=\eta(x,p,t)$. It is well known that the Langevin equations for the Brownian particle is written by the position $x$ and the momentum $p$ as

$$\frac{dx}{dt}=\frac{p}{m}, \quad \frac{dp}{dt}=-\frac{dV(x)}{dx}-\gamma(x,p)\frac{p}{m}+\eta(x,p,t). \tag{1}$$

Usually, the noise is assumed to be Gaussian, with zero-averaged and delta-correlated in time $t$, such that it satisfies,

$$\langle \eta(x,p,t)\rangle=0, \quad \langle \eta(x,p,t)\eta(x,p,t')\rangle=2D(x,p)\delta(t-t'), \tag{2}$$

where the correlation strength of the multiplicative noise, i.e. diffusion coefficient $D(x,p)$, is a function of the variables $x$ and $p$. For such nonlinear Langevin equations (1) and (2), the associated FP equation (in Zwanzig's or backward-Ito's rule [2, 40]) is written [16] as

$$\frac{\partial\rho}{\partial t}=-\frac{p}{m}\frac{\partial\rho}{\partial x}+\frac{\partial}{\partial p}\left[\frac{dV(x)}{dx}+\gamma(x,p)\frac{p}{m}\right]\rho+\frac{\partial}{\partial p}D(x,p)\frac{\partial\rho}{\partial p}, \tag{3}$$

where $\rho\equiv\rho(x,p,t)$ is the probability distribution function. Eq.(3) is formally the Klein-Kramers equation only if the diffusion coefficient is a constant. The FP equation for the overdamped process is the Smoluchowski equation. With the anomalous diffusion and the generalized FDR, we can find the generalized forms of these two equations [16]. The stochastic process for Eq.(3) is a "standard" Brownian motion, which is of course different from that with the time-dependent nonuniform temperature [41] and that with the scaled Brownian motion [42].

For Eq.(3), so far, nothing has been said about requiring that $\rho(x,p,t)$ must approach an equilibrium distribution at long times. If there is not enough friction to dampen the heating effect of the noise, we expect that the system will "run away" so that there is no long time steady state. If there is too much friction for the noise, the system will cool down and "die". In fact, not much is known in general about the long time steady-state solution of such an arbitrary nonlinear FP equation. If a steady-state solution is found, then it implies a fluctuation-dissipation relation (FDR) between $\gamma(x,p)$ and $D(x,p)$ [2]. In the simplest case if the diffusion and the friction coefficients are both constant and satisfy the FDR, i.e. $D/\gamma=kT$ with temperature $T$ and Boltzmann constant $k$, then the FP equation (3) has a long time steady-state solution and it is a Maxwell-Boltzmann distribution. Thus the system reaches a thermal equilibrium state. In a more general case, if the diffusion coefficient and the friction coefficient are both position/momentum-dependent and satisfy a generalized FDR[16], i.e. $D/\gamma\equiv f(E)$, and

$$D(x,p)=\gamma(x,p)\beta^{-1}(1-\kappa\beta E)_+ \tag{4}$$

with $\beta^{-1}\equiv kT$, where $E\equiv V(x)+p^2/2m$ is the energy, $f(E)$ is a continuously differentiable function and the parameter $\kappa$ is defined as $\kappa\equiv -f'(0)=-[\partial f(E)/\partial E]_{E=0}$ which measures the distance away from the MB equilibrium, the long time steady-state solution of the

FP equation (3) exists and it is a power-law distribution [16], given by

$$\rho_s(x,p) = Z_\kappa^{-1} (1-\kappa\beta E)_+^{\frac{1}{\kappa}}, \quad (5)$$

where $Z_\kappa = \iint dxdp (1-\kappa\beta E)_+^{1/\kappa}$ is the normalization constant, and $(z)_+ = z$ for $z > 0$ and is zero otherwise. Eq.(5) is equal to the $q$-distribution in nonextensive statistics. The power-law distribution can be either a stationary nonequilibrium distribution or an equilibrium distribution [43], which depends on the information about specific form of the diffusion coefficient, the existence and the uniqueness of equilibrium etc.

## 3. Time-dependent solution of the FP equation with anomalous diffusion

Here we focus on the time evolution of the FP equation (3). Before finding the time-dependent solution, we make some simplifying assumptions. We let $V(x)=0$ and the friction coefficient be a constant. Accordingly, the diffusion coefficient in the FDR (4) only becomes a function of $p$. In this case, the probability distribution is a function of $(p, t)$, and integrating FP equation (3) over the position $x$ it becomes

$$\frac{\partial \rho(p,t)}{\partial t} = \frac{\gamma}{m}\frac{\partial}{\partial p}[p\rho(p,t)] + \frac{\partial}{\partial p}D(p)\frac{\partial \rho(p,t)}{\partial p}, \quad (6)$$

and the generalized FDR (4) reads

$$D(p) = \gamma kT(1-\kappa\frac{p^2}{2mkT})_+. \quad (7)$$

Eq.(6) is sometimes called the Rayleigh's equation [1] only if the diffusion coefficient is a constant. The momentum-dependent diffusion coefficient which has the form like (7) is known as anomalous diffusion [12], such as the velocity-space diffusion of the plasma in a superthermal radiation field [10], the momentum-space diffusion in an optical lattice [13] and the velocity-space diffusion in nonlinear Brownian motion [4].

After we take the dimensionless variable substitutions, $\tilde{p} = (mkT)^{-\frac{1}{2}} p$ and $\tilde{t} = (\gamma m^{-1})t$, Eqs.(6) and (7) turn to

$$\frac{\partial \rho(\tilde{p},\tilde{t})}{\partial \tilde{t}} = \frac{\partial}{\partial \tilde{p}}[\tilde{p}\rho(\tilde{p},\tilde{t})] + \frac{\partial}{\partial \tilde{p}}D(\tilde{p})\frac{\partial \rho(\tilde{p},\tilde{t})}{\partial \tilde{p}}, \quad (8)$$

with

$$D(\tilde{p}) = \left(1-\kappa\frac{\tilde{p}^2}{2}\right)_+. \quad (9)$$

As usual, the initial condition for the probability distribution is given as

$$\rho(\tilde{p},0) = \delta(\tilde{p}), \quad (10)$$

and the boundary condition is selected as

$$\rho(\pm\infty,\tilde{t}) = 0 \quad \text{for} \quad \kappa < 0, \quad (11)$$

$$\rho(\pm p_{max},\tilde{t}) = 0 \quad \text{for} \quad \kappa > 0. \quad (12)$$

To solve Eq.(8), according to the eigen-function expansion method [1], we assume the distribution function can be expanded as

$$\rho(\tilde{p},\tilde{t}) = \sum_\lambda \varphi_\lambda(\tilde{p}) e^{-\lambda\tilde{t}}, \quad (13)$$

where $\varphi_\lambda(\tilde{p})$ is the eigen-function of the FP operator $L_{FP}$,

$$L_{FP} = \frac{\partial}{\partial \tilde{p}} \tilde{p} + \frac{\partial}{\partial \tilde{p}} D(\tilde{p}) \frac{\partial}{\partial \tilde{p}}, \tag{14}$$

and $\lambda$ is the eigenvalue. In this method, it has been proved that the eigenvalue is always positive [1], i.e. $\lambda \geq 0$. Based on properties of the exponential function in Eq.(13), obviously, the terms with larger eigenvalues will give a relatively small contributions. As a reasonable approximation in the sum, one can omit those terms with larger eigenvalues and only retains the terms with smaller eigenvalues. We find that when the time $t \to \infty$, only the term with eigenvalue $\lambda$=0 remains and the other terms all vanish. Namely,

$$\rho(\tilde{p},\infty) = \varphi_{\lambda=0}(\tilde{p}). \tag{15}$$

Therefore, the eigen-function $\varphi_0(\tilde{p})$ for $\lambda$=0 is exactly the long time steady-state solution of Eq.(8). Substituting Eqs.(9) and (13) into Eq.(8), we have

$$-\sum_\lambda \lambda \varphi_\lambda(\tilde{p}) e^{-\lambda \tilde{t}} = \sum_\lambda e^{-\lambda \tilde{t}} \left[ \varphi_\lambda(\tilde{p}) + (1-\kappa)\tilde{p} \frac{d\varphi_\lambda(\tilde{p})}{d\tilde{p}} + D(\tilde{p}) \frac{d^2\varphi_\lambda(\tilde{p})}{d\tilde{p}^2} \right], \tag{16}$$

For each term in the sums of Eq.(16) we have that

$$\left(1 - \kappa \frac{\tilde{p}^2}{2}\right) \frac{d^2 \varphi_\lambda(\tilde{p})}{d\tilde{p}^2} + (1-\kappa)\tilde{p} \frac{d\varphi_\lambda(\tilde{p})}{d\tilde{p}} + (1+\lambda)\varphi_\lambda(\tilde{p}) = 0. \tag{17}$$

This equation can be written as the associated Legendre differential equation of order $l$,

$$(1-y^2) \frac{d^2 \phi_\lambda(y)}{dy^2} - 2y \frac{d\phi_\lambda(y)}{dy} + \left[l(l+1) - \frac{\mu^2}{1-y^2}\right] \phi_\lambda(y) = 0, \tag{18}$$

by using the variable transformations (19)-(22) as follows,

$$y = \sqrt{\frac{\kappa}{2}} \tilde{p}, \tag{19}$$

$$\phi_\lambda(y) = (1-y^2)^{\frac{\mu}{2}} \varphi_\lambda(\tilde{p}), \tag{20}$$

$$l = \frac{-\kappa \pm \sqrt{(\kappa+2)^2 + 8\lambda\kappa}}{2\kappa}, \tag{21}$$

$$\mu = -\kappa^{-1}. \tag{22}$$

We notice that the plus-minus sign in Eq.(21) does not change $l(l+1)$ and hereby does not change the mathematical form of Eq.(18), which means we do not need to treat the plus-minus sign in Eq.(21) as two different cases separately. Accordingly, the solution of Eq.(17) can be found by Eq.(18) [44] as

$$\varphi_\lambda(\tilde{p}) = C_\lambda \sqrt{\frac{\kappa}{2}} \left(1 - \kappa \frac{\tilde{p}^2}{2}\right)_+^{\frac{1}{2\kappa}} P_l^{-\frac{1}{\kappa}}\left(\sqrt{\frac{\kappa}{2}} \tilde{p}\right), \tag{23}$$

where $P_l^\mu(x)$ is the associated Legendre function of the first kind. Substituting Eq.(23) into Eq.(13), we finally write the time-dependent solution of Eq.(8),

$$\rho(\tilde{p},\tilde{t}) = \sum_\lambda C_\lambda \sqrt{\frac{\kappa}{2}} \left(1 - \kappa \frac{\tilde{p}^2}{2}\right)_+^{\frac{1}{2\kappa}} P_l^{-\frac{1}{\kappa}}\left(\sqrt{\frac{\kappa}{2}}\tilde{p}\right) e^{-\lambda \tilde{t}}. \qquad (24)$$

For $\kappa>0$, with the aid of the initial condition (10) as well as the orthogonal relation between the associated Legendre functions,

$$\int_{-1}^{1} P_l^\mu(z) P_{l'}^\mu(z) dz = 0, \quad l \neq l', \qquad (25)$$

the factor $C_\lambda$ can be determined by the formula

$$C_\lambda = \frac{P_l^\mu(0)}{\int_{-1}^{1}\left[P_l^\mu(z)\right]^2 dz}. \qquad (26)$$

If $1/\kappa$ is a positive integer, the eigenvalue of the associated Legendre equation (18) can be further calculated [45] by

$$l = \frac{1}{\kappa}, \ \frac{1}{\kappa}+1, \ \frac{1}{\kappa}+2,\ldots. \qquad (27)$$

And thus the eigenvalue $\lambda$ can be derived using Eq.(21) by

$$\lambda = \frac{\kappa^2 l(l+1)-(1+\kappa)}{2\kappa}. \qquad (28)$$

where $l=1/\kappa$ corresponds to $\lambda=0$.

At the long time, the solution (24) approaches to a stationary $\rho(\tilde{p},\infty)=\varphi_0(\tilde{p})$, which is found to be a power-law distribution given in (5) (see the Appendix), i.e.,

$$\rho(\tilde{p},\infty) = A\left(1 - \kappa \frac{\tilde{p}^2}{2}\right)^{\frac{1}{\kappa}}, \ \text{with} \ A = \sqrt{\frac{\kappa}{2\pi}} \frac{\Gamma(\kappa^{-1}+\frac{3}{2})}{\Gamma(\kappa^{-1}+1)}. \qquad (29)$$

## 4. Numerical test and comparisons

4.1. *General Test*

The precise time-evolution analytical solution of FP equation (6) with the FDR (7) is expressed by the sum in (24). In order to examine the validity of the analytical solution, we employ the implicit Runge-Kutta method [46] to do numerical studies of the solution of Eq.(6) with the FDR (7) for $\kappa>0$, and then compare with our analytical solution (24). Because we can only take the first finite terms in the infinite terms of (24) to calculate the analytical solution (here we have taken the first 30 terms in (24) as the analytical results), its initial function is also finite instead of the delta function. Thus, the initial condition for the analytical results is set as $\rho(p,0)$ which can be calculated by (24). We set the same initial function as $\rho(p,0)$ for the numerical studies. The boundary condition (12) is taken the same for both the analytical and numerical studies.

In principle, the calculation results of a two-variable function $\rho(p,t)$ can be shown using 3D graphics. But because the 3D graphics are not good to clearly observe the differences between two results in the analytical and numerical methods, we here drew 2D graphics of the probability distribution functions at a fixed time or fixed momentum. For example, the momentum distribution $\rho(p)$ is shown in Fig.1 for time $t=0.1$ and $t=1$, respectively, and the time evolution $\rho(0,t)$ is shown in Fig.2 at the momentum $p=0$. Two types of graphics are given respectively for $\rho(p)$ and $\rho(0,t)$ and for the parameter $\kappa = 0.3$ and $0.5$.

In Figs.1-2, we observe that the analytical results are all exactly the same as the numerical results, which therefore have demonstrated the accuracy and validity of the time-dependent analytical solution (24) of the Fokker-Planck equation (6) with the anomalous diffusion described by (7).

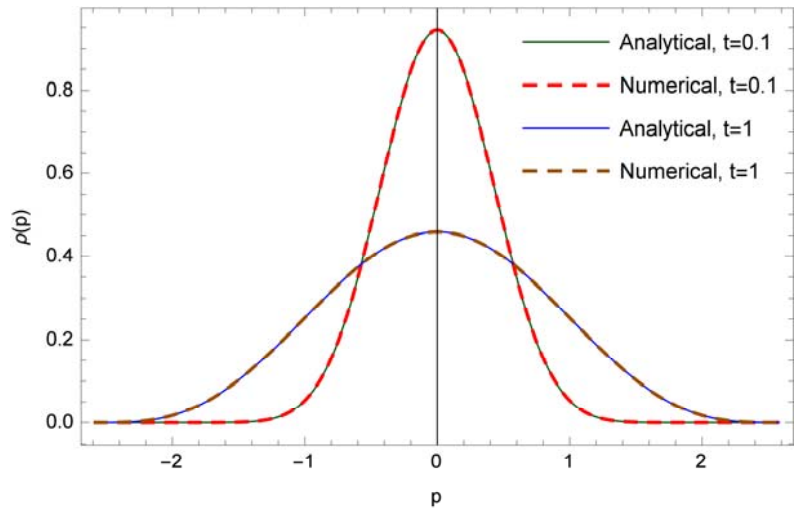

(a) for $\kappa = 0.3$

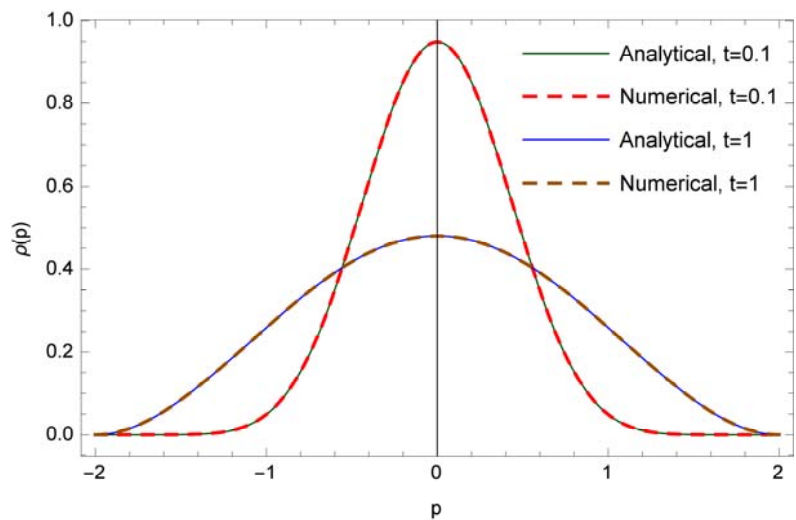

(b) for $\kappa = 0.5$

Fig. 1. The momentum distribution at $t=0.1$ and $t=1$ for $\kappa=0.3$ (a) and $\kappa=0.5$ (b), respectively.

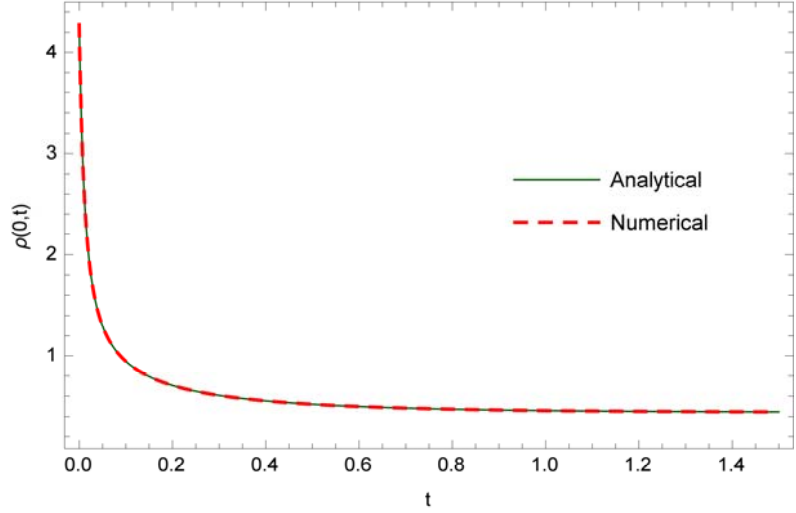

(a) $\kappa = 0.3$

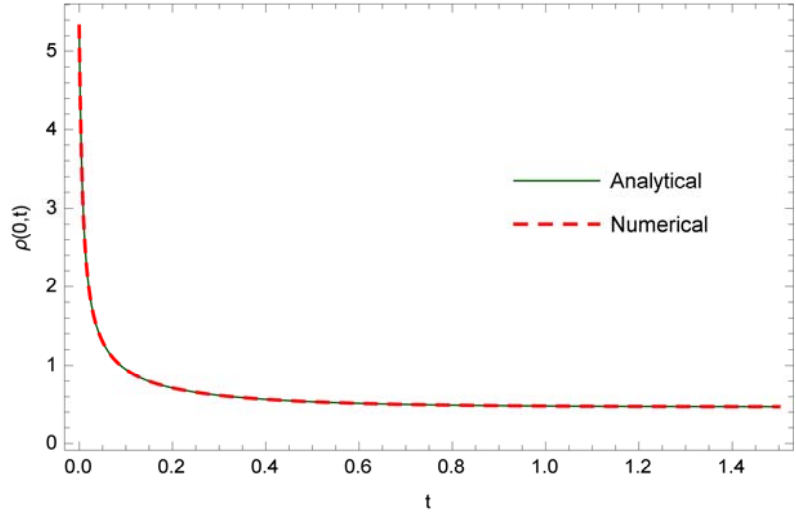

(b) $\kappa = 0.5$

Fig.2. The time evolution of the distribution function at $p=0$ for $\kappa=0.3$ (a) and $\kappa=0.5$ (b), respectively.

4.2. *Application to Ornstein-Uhlenbeck process.*

The Ornstein-Uhlenbeck process describes a Brownian particle moving in the media with constant friction coefficient $\gamma$. Usually, the diffusion coefficient $D$ used to be a constant (normal diffusion). Consequently, the corresponding FP equation [1] reads

$$\frac{\partial f(p,t)}{\partial t} = \frac{\gamma}{m}\frac{\partial}{\partial p}[pf(p,t)] + D\frac{\partial^2 f(p,t)}{\partial p^2}, \qquad (30)$$

where $f(p,t)$ is the time-dependent momentum distribution function. If the initial condition is selected a delta function, i.e. $f(p,0)=\delta(p)$, then the time-dependent solution of the FP equation (30) is exactly,

$$f(p,t) = \sqrt{\frac{\gamma}{2\pi mD(1-e^{-2\gamma t})}} \exp\left[-\frac{\gamma p^2}{2mD(1-e^{-2\gamma t})}\right], \qquad (31)$$

which at long time tends to the Gaussian distribution. The FP equation (6) as well as Eq.(7) can be regarded as a generalization of the FP equation for Ornstein-Uhlenbeck process with the anomalous diffusion, where the diffusion coefficient $D$ is given by Eq.(7) as a momentum-dependent function. In this situation, the time-dependent solution is exactly replaced by (24).

Numerically, we can make comparison between the solution (31) with the normal diffusion and our results (24) with the anomalous diffusion, also including Eqs.(26)-(28). The calculations of Eq.(31) can be made taking $\gamma=D=1$. The calculation results for the comparison between (31) and (24) are shown in Fig.3.

We show that in the beginning time ($t = 0.005\sim 0.2$ in Fig.3. (a)) the relaxation of the probability distributions for the two situations, the normal diffusion and the anomalous diffusion, are very close to each other. With the evolution of the time ($t=$ 0.2~2.5 in Fig.3.(b)), the two situations separate gradually and the difference becomes more and more obvious, and finally the distribution with normal diffusion tends to the Gaussian distribution at an equilibrium state but that with the anomalous diffusion approaches to the stationary power-law distribution given by (29).

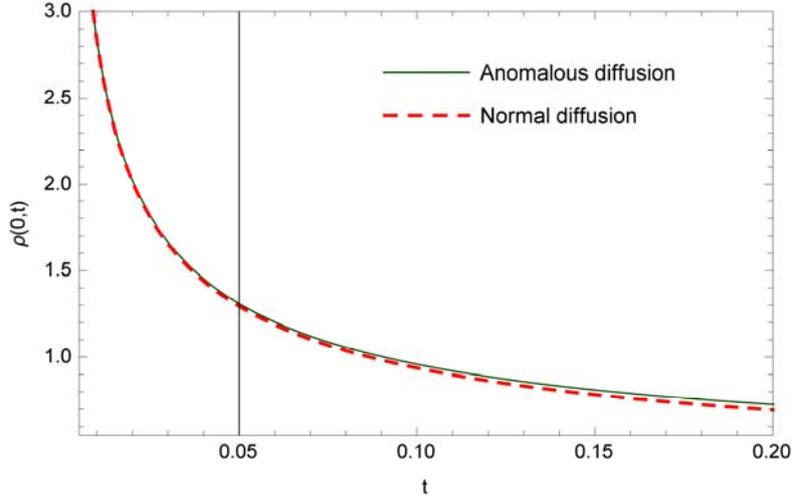

(a) The time $t = 0.005 \sim 0.2$.

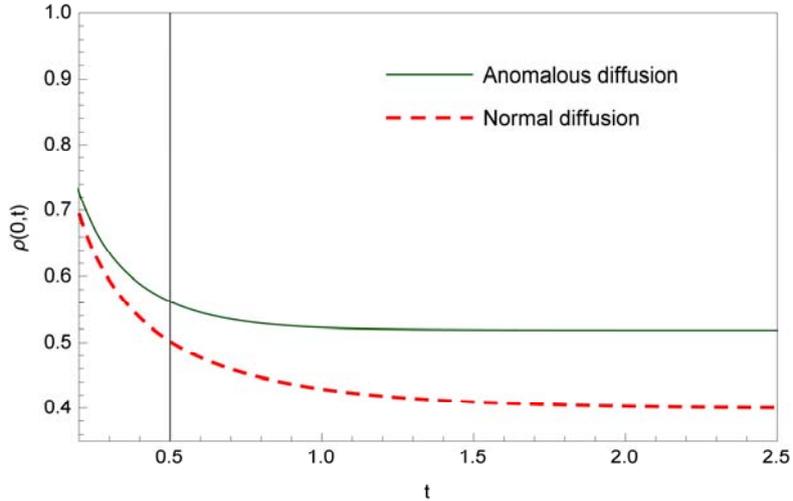

(b) The time $t= 0.2 \sim 2.5$.

Fig. 3. The time-evolution of the Fokker-Planck equation for the Ornstein-Uhlenbeck process with normal diffusion and with the anomalous diffusion for $\kappa$=0.9.

## 5. Conclusion

Many different forms of the stationary power-law distributions can be generated exactly from the well-known Langevin equation of the Brownian motion and the associated Fokker-Planck (FP) equations under a generalized fluctuation-dissipation relation (FDR), but the time behavior of the equations is still unknown. In this work, we study the time evolution of the probability distribution from the standard FP equation (in Zwanzig's or backward-Ito's rule) based on the Langevin equation of the Brownian motion with an anomalous diffusion in a complex medium. The diffusion is in the momentum space and the diffusion coefficient is given as a function of the momentum $p$ in (7) under a generalized FDR, i.e. $D(p)=\gamma kT[1-\kappa p^2/(2mkT)]$. We exactly obtained the time-dependent analytical solution of the FP equation (6) with the anomalous diffusion. The precise time-dependent solution is given by the series form (24), which includes the associated Legendre function, and the eigenvalues are determined by Eqs.(27)-(28) for $\kappa>0$. At long time, the time-dependent solution approaches to the stationary power-law distribution in nonextensive statistics.

As a general test, we employed Runge-Kutta method to do numerical studies of the time-dependent FP equation for $\kappa>0$, and then compared with our analytical solution. The numerical calculations in Figs.1-2 showed excellent accordance with the analytical solution and thus demonstrated the accuracy and validity of the solution. As an application test, numerically we illustrated the timer-dependent solutions of the FP equations for the Ornstein-Uhlenbeck process with normal diffusion and anomalous diffusion, respectively, and made the comparison. In Fig.3 we showed that at the beginning time (in Fig.3. (a)) the relaxations of two probability distributions are very close to each other, with the evolution of time (in Fig.3.(b)) the difference is more and more obvious, and finally at long time the probability distribution with normal diffusion tends to the stationary Gaussian distribution but that with the anomalous diffusion approaches to the stationary power-law distribution.

**Acknowledgments**

This work is supported by the National Natural Science Foundation of China under grant No 11175128 and by the Higher School Specialized Research Fund for Doctoral Program under grant No 20110032110058.

**Appendix**

If we take $t\to\infty$, the long time stationary solution in (24) is

$$\rho(\tilde{p},\infty) = C_0 \sqrt{\frac{\kappa}{2}} \left(1-\kappa\frac{\tilde{p}^2}{2}\right)_+^{\frac{1}{2\kappa}} P_{1/\kappa}^{-1/\kappa}\left(\sqrt{\frac{\kappa}{2}}\tilde{p}\right). \tag{A1}$$

For $\kappa>0$, according to the associated Legendre function given in [44] by

$$P_{\frac{1}{\kappa}}^{-\frac{1}{\kappa}}\left(\sqrt{\frac{\kappa}{2}}\tilde{p}\right) = \frac{2^{-\frac{1}{\kappa}}}{\Gamma(\kappa^{-1}+1)}\left(1-\kappa\frac{\tilde{p}^2}{2}\right)^{\frac{1}{2\kappa}},$$

we can write the stationary solution (A1) as

$$\rho(\tilde{p},\infty) = A\left(1-\kappa\frac{\tilde{p}^2}{2}\right)^{\frac{1}{\kappa}}, \qquad (A2)$$

where the factor reads

$$A = C_0 \sqrt{\frac{\kappa}{2}} \frac{2^{-\frac{1}{\kappa}}}{\Gamma(\kappa^{-1}+1)}. \qquad (A3)$$

If $1/\kappa$ is a positive integer, further one has

$$C_0 = \frac{P_{\frac{1}{\kappa}}^{-\frac{1}{\kappa}}(0)}{\int_{-1}^{1}\left[P_{\frac{1}{\kappa}}^{-\frac{1}{\kappa}}(x)\right]^2 dx} = \left(\frac{1}{\kappa}+\frac{1}{2}\right)\frac{(\kappa^{-1})!}{2^{\frac{1}{\kappa}}\Gamma(\kappa^{-1}+1)} = \left(\frac{1}{\kappa}+\frac{1}{2}\right)\frac{2\Gamma(2\kappa^{-1})}{2^{\frac{1}{\kappa}}\Gamma(\kappa^{-1})}. \qquad (A4)$$

Due to the relation [44],

$$\frac{\Gamma(2\kappa^{-1})}{\Gamma(\kappa^{-1})} = \frac{1}{\sqrt{\pi}} 2^{\frac{2}{\kappa}-1} \Gamma\left(\frac{1}{\kappa}+\frac{1}{2}\right),$$

the factor $C_0$ is simplified as

$$C_0 = \frac{1}{\sqrt{\pi}} 2^{1/\kappa} \Gamma\left(\frac{1}{\kappa}+\frac{3}{2}\right).$$

And then the factor (A3) becomes

$$A = \sqrt{\frac{\kappa}{2\pi}} \frac{\Gamma(\kappa^{-1}+\frac{3}{2})}{\Gamma(\kappa^{-1}+1)}. \qquad (A5)$$

It is worth to point that the factor is exactly the reciprocal normalization in (5),

$$Z_\kappa^{-1} = \left[\int_{-\sqrt{2/\kappa}}^{\sqrt{2/\kappa}} \left(1-\kappa\frac{\tilde{p}^2}{2}\right)^{\frac{1}{\kappa}} d\tilde{p}\right]^{-1} = A. \qquad (A6)$$

**Reference**

[1] H. Risken, *The Fokker–Planck Equation: Methods of Solution and Applications*, 2nd Edition, Springer, Berlin, 1989; N. G. van Kampen, *Stochastic Processes in Physics and Chemistry*, North-Holland, Amsterdam, 1981.
[2] R. Zwanzig, *Nonequilibrium Statistical Mechanics*, Oxford University Press, New York, 2001.
[3] R.Vielen, *Astron. Astrophys.* **60** (1977) 263.
[4] B, Lindner, *New J.Phys.* **9** (2007) 136.
[5] A. G. Cherstvy, A. V. Chechkin and R. Metzler, *New J.Phys.* **15** (2013) 083039.
[6] A. G. Cherstvya and R. Metzler, *Phys. Chem. Chem. Phys.* **15** (2013) 20220.
[7] A. G. Cherstvy, A. V. Chechkin and R. Metzler, *Softer Matter* **10** (2014) 1591.
[8] T. Kühn, T. Ihalainen, J. Hyväluoma, N. Dross, S. Willman, J. Langowski, M. Vihinen-Ranta, J. Timonen, *PLoS ONE* **6** (2011) e22962.
[9] R. Metzler, J. -H. Jeon, A. G. Cherstvy and E. Barkai, *Phys.Chem.Chem.Phys.* **16** (2014) 24128.
[10] A. Hasegawa, K. Mima, and M. Duong-van, *Phys. Rev. Lett.* **54** (1985) 2608.
[11] G. Kaniadakis and P. Quarati, *Physica A* **237** (1997) 229.
[12] G. Kaniadakis and G. Lapenta, *Phys. Rev. E* **62** (2000) 3246.
[13] E. Lutz, *Phys. Rev. A* **67**(R) (2003) 051402.
[14] B. Liu and J. Goree, *Phys. Rev. Lett.* **100** (2008) 055003.
[15] B. Liu, J. Goree and Y. Feng, *Phys.Rev.E* **78** (2008) 046403.
[16] J. L. Du, *J. Stat. Mech.* (2012) P02006.
[17] R. Guo and J. L. Du, *J. Stat. Mech.* (2013) P02015.
[18] C. Tsallis and D. J. Bukman, *Phys.Rev.E* **54**(R) (1996) 2197.
[19] L. Borland, *Phys.Rev.E* **57** (1998) 6634.

[20] E. K. Lenzi, C. Anteneodo and L. Borland, *Phys.Rev.E* **63** (2001) 051109.
[21] E. M. F. Curado and F. D. Nobre, *Phys.Rev.E* **67** (2003) 021107.
[22] E. K.Lenzi, L.C. Malacarne, R.S. Mendes, I.T. Pedron, *Physica A* **319** (2003) 245.
[23] J. S. Andrade, G. F. T. daSilva, A. A. Moreira, F. D. Nobre and E. M. F. Curado, *Phys. Rev. Lett.* **105** (2010) 260601.
[24] C. Tsallis, *Introduction to Nonextensive Statistical Mechanics: Approaching a Complex World*, Springer, New York, 2009.
[25] M. P. Leubner, *Astrophys. J.* **604** (2004) 469.
[26] T. Kronberger, M.P. Leubner and E. van Kampen, *Astron. Astrophys.* **453** (2006) 21.
[27] J. L. Du, *Europhys. Lett.* **75** (2006) 861.
[28] V. F. Cardone, M.P. Leubner, A. Del Popolo, *Mon. Not. R. Astron. Soc.* **414** (2011) 2265.
[29] Y. H. Zheng, *Europhys. Lett.* **102** (2013) 10009.
[30] V. M. Vasyliunas, *J. Gerophys. Res.* **73** (1968) 2839.
[31] M. P. Leubner, *Phys. Plasmas* **11** (2004) 1308.
[32] Z. P. Liu, L.Y. Liu and J. L. Du, *Phys. Plasmas* **16** (2009) 072111.
[33] J. Y. Gong, Z. P. Liu and J. L. Du, *Phys. Plasmas* **19** (2012) 093706.
[34] J. L. Du, *Phys. Plasmas* **20** (2013) 092901 and the references therein; H.N.Yu and J.L.Du, *Ann. Phys.* **350** (2014) 302.
[35] J. R. Claycomb, D. Nawarathna, V. Vajrala, J.H. Miller, *J. Chem. Phys.* **121** (2004) 12428.
[36] V. Aquilanti, K.C. Mundim, M. Elango, S. Kleijn, T. Kasai, *Chem. Phys. Lett.* **498** (2010) 209.
[37] J. L. Du, *Physica A* **391** (2012) 1718.
[38] C. Yin, J.L. Du, *Physica A* **395** (2014) 416; *Physica A* **407** (2014) 119.
[39] C. Yin, R. Guo and J.L. Du, *Physica A* **408** (2014) 85.
[40] J. Dunkel and P. Hänggi, *Phys. Rep.* **471** (2009) 1.
[41] A. Fuliński, *J. Chem. Phys.* **138** (2013) 021101.
[42] J. -H. Jeon, A. V. Chechkin and R. Metzler, *Phys.Chem.Chem.Phys.* **16** (2014) 15811.
[43] R. Guo and J. L. Du, *Physica A* **406** (2014) 281.
[44] M. Abramowitz, I.A. Stegun, *Handbook of Mathematical Functions with Formulas, Graphs and Mathematical Tables*, National Bureau of Standards Applied Mathematics Series, Vol. 55, US Government Printing Office, Washington DC, 1972.
[45] R. Courant and D. Hilbert, *Methods of Mathematical Physics*, vol. 1, John Wiley & Sons, Inc., New York, 1953
[46] K. Atkinson, W. Han and D. Stewart, *Numerical solution of ordinary differential equations*, John Wiley & Sons, Inc., New Jersey, 2009.